# DEIMoS: an open-source tool for processing high-dimensional mass spectrometry data

Sean M. Colby, Christine H. Chang, Jessica L. Bade, Jamie R. Nunez, Madison R. Blumer, Daniel J. Orton, Kent J. Bloodsworth, Ernesto S. Nakayasu, Richard D. Smith, Yehia M. Ibrahim, Ryan S. Renslow,* Thomas O. Metz*

*Corresponding authors

Biological Sciences Division
Pacific Northwest National Laboratory
902 Battelle Boulevard
P. O. Box 999, MSIN K8-98
Richland, WA, 99352 USA

**Abstract**

We present DEIMoS: Data Extraction for Integrated Multidimensional Spectrometry, a Python application programming interface (API) and command-line tool for high-dimensional mass spectrometry data analysis workflows that offers ease of development and access to efficient algorithmic implementations. Functionality includes feature detection, feature alignment, collision cross section (CCS) calibration, isotope detection, and MS/MS spectral deconvolution, with the output comprising detected features aligned across study samples and characterized by mass, CCS, tandem mass spectra, and isotopic signature. Notably, DEIMoS operates on N-dimensional data, largely agnostic to acquisition instrumentation; algorithm implementations simultaneously utilize all dimensions to (i) offer greater separation between features, thus improving detection sensitivity, (ii) increase alignment/feature matching confidence among datasets, and (iii) mitigate convolution artifacts in tandem mass spectra. We demonstrate DEIMoS with LC-IMS-MS/MS data to illustrate the advantages of a multidimensional approach in each data processing step.

**Introduction**

The ability to process raw instrument data reliably and accurately is critical to any molecular profiling assay. Though useful, commercial software solutions provided by vendors of mass spectrometry (MS) instrumentation lack flexibility required to rapidly adapt to evolving community needs. Demand for open-source, community-driven development has motivated researchers to pursue alternatives across instrument platforms, for example liquid or gas chromatography (LC or GC) and ion mobility spectrometry (IMS) coupled to mass spectrometry (MS), including tandem mass spectrometry (MS/MS). Software implementations also differ in their offered functionality: data input/output, multidimensional feature detection, alignment across samples, isotope detection, and deconvolution of MS/MS spectra. However, few available open-source, platform-agnostic solutions provide such core functionality for data of high-dimensionality, hindering the development and application of new instrumentation and analysis paradigms.

These limitations are predominantly tied to the specificity—and thus, relative inflexibility—of existing software and algorithm implementations. For example, LC or GC coupled to MS or MS/MS results in two primary feature dimensions, the retention time/index and MS mass-to-charge ratio (*m/z*), which is reflected in community software algorithms[1–13]. Existing feature detection algorithms are tailored to the underlying data: features are detected in one or two dimensions, and dimensions are often inflexibly constrained based on respective assumptions. In the short term, instrumentation advances[14–19] force platform-specific software to either ignore additional dimensions—for instance, summing across the least-distinguishing dimensions—or iteratively apply one- or two- dimensional algorithms[20]. Over time, wider instrument adoption engenders development to extend or modify existing algorithms to take full advantage of additional separation dimensions[20,21]. The problem is thus cyclical in nature: for software to adapt to technology, the technology must be mature and widely used, but for technology to mature and achieve widespread adoption, instrument data must be robustly analyzed by user-friendly software.

To overcome this paradox, instrument vendors have historically developed and supplied the software required to process the data, e.g. Agilent MassHunter, Bruker MetaboScape, and Waters Progenesis QI. However, vendor offerings have their own limitations[22]. Because the underlying software is proprietary, details of underlying algorithm implementations are neither available to the public via open-source codebases nor sufficiently documented in publications. Moreover, vendor software is often tailored to a specific instrument and involves proprietary data formats, limiting one-to-one comparison of data across different instrument or vendor types. In some cases, existing software can also lack customizability; for instance, algorithm selection tends to be fixed to specific peak detection, alignment, and deconvolution implementations, unless additional options are explicitly implemented by the vendor. Thus, users are subject to the functionality provided by the vendor software, or must assemble multiple software solutions into one workflow.[3,23] Finally, many vendor solutions are automated only to a small degree, thus impeding reproducibility, and are not amenable to high-performance (HPC) or cloud computing.

As a result, the metabolomics community has worked to develop open-source solutions. Notable contributions include MZmine[8], ProteoWizard[24], MS-DIAL[20], XCMS[25], OpenMS[10], Maven[26], MetaboAnalyst[27], and more. Each cover either some or all of the steps in a typical metabolomics workflow and are positioned to analyze GC-MS and/or LC-MS data (MS-DIAL additionally handles some aspects of LC-IMS-MS data) and tandem MS. These software tools offer insight into best practices and algorithm implementations and serve as foundational references for our work. However, challenges remain in supporting data of arbitrary dimensionality, generalizing algorithmic implementations to operate in native dimensionality, offering flexibility and control over the analysis workflow, and scaling efficiently to computational resources.

To this end, we present the design and implementation of DEIMoS, or Data Extraction for Multidimensional Spectrometry, and include an initial evaluation on LC-IMS-MS/MS data from analysis of blood plasma samples. DEIMoS's functionality is generalized through use of N-dimensional signal processing algorithms from the open-source, efficient, and widely used Python-based scientific computing packages NumPy[28] and SciPy.[29] Additionally, DEIMoS's design makes minimal assumptions about each underlying dimension. As a result, researchers may analyze GC-MS, LC-MS, IMS-MS, or LC-IMS-MS data, or another hypothetical MS-based platform, with or without MS/MS, using the same software with minimal reconfiguration. The underlying source code has also been written to account for hypothetical additional separation or analytical dimensions that may be introduced as instrumentation continues to advance (e.g., solid phase extraction[30] and associated chemical class-based separation, cryogenic infrared spectroscopy[31], or multiplexed higher resolution ion mobility separations such as provided by structures for lossless ion manipulations, SLIM[32]). That is, calls to the application programming interface (API) and logic of the analysis may change, but the underlying source code can remain intact. This paradigm facilitates rapid advancement in metabolomics and introduces the potential to unify community efforts in informatics software development.

Furthermore, DEIMoS benefits from Python's rich existing ecosystem for scientific programming and offers even greater flexibility beyond the core API. DEIMoS's functionality is organized into several modules, each addressing one or more key data processing steps, including file input and output, peak detection, alignment, isotope detection, MS2 spectra extraction and deconvolution, and data subsetting operations. We describe each module relative to LC-IMS-MS/MS data, which represents higher dimensionality – and, by extension, complexity – among most current metabolomics analysis techniques. Data acquired on other platforms – e.g. LC-MS(/MS), GC-MS(/MS) – require similar processing, but in a lower dimensional space. Future algorithms and additional dimensions of data may be slotted in easily.

## Methods

### Experimental methods

To demonstrate DEIMoS, we examined LC-IMS-MS/MS data from a large study of human plasma samples consisting of 40 quality control (QC) samples from the NIST Standard Reference Material 1950[33] and 112 study samples. An internal standard mixture consisting of D4-malonic acid, D4-succinic acid, D5-glycine, D4-citric acid, $^{13}$C6-fructose, D5-L-tryptophan, D4-lysine, D7-alanine, D35-stearic acid, D5-benzoic acid, and D15-octanoic acid was added prior to extraction. Each sample was spiked with 50 µL of a solution of the internal standards at 0.166 mg/mL in water. Metabolites and lipids were extracted with concomitant protein precipitation using the Matyash protocol[34] described previously[35]. The metabolite layer was removed and dried *in vacuo*. Lipid and protein layers were not analyzed.

An Agilent 1260 Infinity II high flow liquid chromatography system (San Jose, CA) equipped with a Vial Sampler and Binary Pump was used to inject and chromatographically separate samples prior to introduction to the ion mobility spectrometry-mass spectrometry instrument. A steady flowrate of 0.300 mL/min was delivered through a Millipore-Sigma (Burlington, MA) SeQuant Zic-pHILIC column (15 cm length, 2.1 mm inner diameter, packed with 5 µm particles). A corresponding guard column of the same packing material was also used. Mobile phases consisted of (A) 20 µM ammonium acetate in water and (B) 100% acetonitrile with the following gradient profile (min, %B): 0, 90; 4, 90; 12, 20; 13, 10; 15, 10; 17, 90. An Agilent 1100 Column Heater was used with a static temperature of 45°C.

Ion mobility spectrometry-tandem mass spectrometry analysis was performed using an Agilent 6560 Ion Mobility LC/Q-TOF system. Spectra were acquired separately in both positive and negative ionization modes. Data were collected in the mass range of 50-1700 *m/z*. Ionization was accomplished using a Dual AJS ESI source, with gas temperature set to 325°C, drying gas set to 5 L/min, nebulizer set to 30 psi, sheath gas temperature set to 275°C, sheath gas flow of 11 L/min, VCap set to 2500 V, nozzle voltage set to 2000 V, and the fragmentor set to 400 V. For the ion mobility separations, the trap fill time was 30,000 µs and released for 300 µs. Frame rate was 1 frame/sec, 19 IM transitions/frame and max drift time set to 50 ms. Fixed collision energies were employed at 10, 20, and 40 eV on alternating frames. Data were collected for 22 minutes immediately following the injection of the sample.

### Overview

DEIMoS was developed as an instrument-independent, high-dimensional metabolomics data analysis tool, and design choices reflect this philosophy. DEIMoS is written in Python, prioritizing user productivity and ease of development and use. While high-level interpreted languages such as Python often suffer from reduced computational efficiency, many popular scientific Python libraries, such as NumPy[28] and SciPy[29], wrap C or Fortran code and are thus highly optimized. In addition, Python is ubiquitous across the sciences and in industry, user-friendly, and largely agnostic to client platform (Windows, macOS, Linux). As a result, Python boasts a large, active, and continually growing community in the sciences, positioning Python-based software for wide adoption both by users and collaborative developers[36]. DEIMoS is one of few Python-based offerings for metabolomics data processing uniquely offering support for data of arbitrary dimensionality, algorithmic implementations that operate in native dimensionality, flexibility and control over the analysis workflow, and efficient scaling to computational resources.

We also recognize that data acquisitions of high dimensionality result in greater memory and processing demands. A typical LC-IMS-MS/MS experiment produces a three-dimensional grid of data with $5 \times 10^{10}$ cells, amounting to approximately 200 GB of memory usage per MS level for 32-bit intensities. DEIMoS's flexible design was written to optimize the efficiency of key memory-intensive algorithms both in cases of memory abundance as well as lower-resource systems. When processing multiple datasets, DEIMoS uses the Snakemake workflow management system[37] to deploy parallel DEIMoS instances to HPC clusters or cloud environments, enabling rapid, simultaneous processing.

We architected DEIMoS to adhere to software development best practices[38], including installation through Anaconda[39] or PyPI[40], in-line documentation via docstrings and aggregation via *Sphinx*[41], unit test implementations with *pytest*[42] coupled with continuous integration and static code coverage analysis, and version control with *Git*[43]. DEIMoS is open-source and freely available online at github.com/pnnl/deimos, and community contributions via pull request are welcome.

<u>File input/output</u>

To accommodate disparate instrument types and manufacturers (e.g. Bruker, Waters, Thermo, Agilent), DEIMoS operates under the assumption that input data are in an open, standard format. As of this publication, the accepted file format for DEIMoS is mzML[44], which contains metadata, separation, and spectrometry data that reproduce the contents of vendor formats. Conversion to mzML from several other formats can be performed using the free and open-source ProteoWizard *msconvert* utility[24].

For optimal use with DEIMoS, we recommend certain *msconvert* options to ensure input data replicate the vendor format as closely as possible (i.e. *msconvert.exe {filename}.{ext} -32 -z -g -outfile {filename}.mzML.gz*). Provided data in mzML format, DEIMoS parses the file contents to build a schema represented internally as a *pandas*[45,46] data frame containing arrays for each separation dimension (e.g. for LC-IMS-MS/MS: retention time, drift time, and *m/z*) and intensity. Because parsing an mzML file can take significant time for large datasets, DEIMoS exports the lightweight, data frame-based representation into the more compact Hierarchical Data Format version 5 (HDF5) file format[47] for subsequent steps. DEIMoS includes adapters to support exporting to CSV, MGF, and mzML for downstream use with other tools (e.g. MAME[48], LIQUID[49], GNPS[50]).

<u>Feature detection</u>

Feature detection, also referred to as peak detection, is the process by which local maxima fulfilling certain criteria (such as sufficient signal-to-noise ratio) are located in the signal acquired by a given analytical instrument. This process results in "features" associated with the analysis of molecular analytes from the sample under study or from chemical, instrument, or random noise. Typically, feature detection involves a mass dimension (*m/z)* as well as one or more separation dimensions, the latter offering distinction among isobaric/isotopic features.

DEIMoS implements an N-dimensional maximum filter from *scipy.ndimage* that convolves the instrument signal with a structuring element, also known as a kernel, and compares the result against the input array to identify local maxima as candidate features or peaks (**Figure 1**). We discuss what qualifies as a dimension in the SI. Additional filters, including integral and average intensity, kurtosis, skew, etc., can be applied to yield statistics for later downselection. To provide additional confidence in detected features, we required that a given feature be observed across analytical triplicates.

Key to this process is the selection of kernel size, which can vary by instrument, dataset, and even compound. For example, in LC-IMS-MS/MS data, peak width increases with increasing *m/z* and drift time, and also varies in retention time. Ideally, the kernel would be the same size as the N-dimensional peak (i.e. wavelets[1,5,12,51]), though computational efficiency considerations in high-dimensional space currently limit the ability to dynamically adjust kernel size. Thus, the selected kernel size should be representative of likely features of interest. In some scenarios, dynamic kernel size may be appropriate, per the kernel scaling discussion in the SI.

The feature detection process is computationally efficient for $N < 3$, but memory-intensive for higher-dimensional data. The data are initially stored in coordinate format, a sparse representation of an N-dimensional array, but must be converted to a dense array to support processing by convolution. To ameliorate memory limitations, partitioning functionality was implemented, which is further discussed in the SI.

For our example data, we determined kernel size for each dimension in two steps. First, we used a single feature of high intensity and well-defined peak shape in each dimension to define parameters for initial feature detection. The footprint of the high-intensity feature, approximately 3-sigma of a Gaussian distribution, was used to determine kernel size relative to the resolution of the underlying data. The kernel was then applied to the full dataset for rough feature coordinate extraction. Second, we sampled the features resulting from step (1) to span each dimension and produced peak statistics as a function of *m/z*. We found that peak width increases in both *m/z* and drift time dimensions with increasing *m/z*, but retention time remains largely invariant. Sampled peak statistics were used to inform final kernel size selection. See **Figure S1** for visualization of peak size analysis.

While we recommend processing the data in its native dimensionality, DEIMoS's algorithms are flexible and can detect features in iterative subspaces, for example 2D followed by 1D, 1D followed by 2D, or successive 1D. We used the same parameters per dimension to evaluate feature detection in all dimensional permutations for LC-IMS-MS data, as shown in **Figures S2 and S3**, to demonstrate the high-level differences between each approach. Features were only kept if they appeared across all three analytical replicates. To compare methods, we (i) compared feature coordinates directly and (ii) used tolerances of ±20 ppm, ±1.5%, and ±0.3 minutes for *m/z*, drift time, and retention time, respectively, based on peak dimensions determined during kernel size selection. We used relative tolerances, such as parts-per-million and percent for *m/z* and drift time, because unlike in the retention time dimension, peak widths in *m/z* and drift time varied with mass (**Figure S1)**. This analysis was performed for all samples from the study, averaged per ionization mode.

Alignment

Alignment is the process by which feature coordinates across samples are adjusted to account for instrument variation (drift, calibration, etc.) such that matching features are aligned to adjust for small differences in coordinates. To perform alignment, we first constructed a model for each dimension of a sample by putatively matching detected features against an in-study reference sample, minimizing the residual, and subsequently applying the fit transform. Next, we matched corresponding features across datasets within a user defined tolerance. We refer to the former as "reference-based alignment" and the latter as "cross-sample alignment."

For reference-based alignment, we defined corresponding features between two samples based on minimum distance in the dimension of interest and selected tolerances to accommodate potentially complex nonlinear relationships. We suggest visualizing putative matches with multiple tolerance selections. Once features were matched, we modeled the relationship between samples using support vector regression (SVR) as implemented in *scikit-learn*[52]. We examined results from various SVR kernels and found that the linear kernel achieved satisfactory results when the instrument misalignment could be corrected by linear regression, whereas the radial basis function (RBF) kernel was able to account for nonlinear relationships. For instance, the RBF SVR model could match features when samples were run consecutively on a degrading LC column. See **Figures 2** and **S4** for examples of nonlinear and linear alignment scenarios, respectively.

Many existing algorithms and implementations can perform cross-sample alignment[8,13,53]. We initially explored use of a modified version of the "join align" method from MZmine[8], but ultimately arrived at an agglomerative clustering-based approach. Though similar with respect to resulting alignment, the agglomerative clustering-based approach was more amenable to processing many samples simultaneously.

Agglomerative clustering is implemented via *scikit-learn* using a custom distance matrix to ensure the maximum linkage distance does not exceed the user-specified tolerance in any one dimension, i.e. Chebyshev distance. Cluster affinity is defined by complete linkage, which uses the maximum of the distances between all observations of two sets to qualify a merge. To ensure that features are merged into clusters across datasets, not within, a connectivity matrix is automatically generated to mask intra-sample linkages. However, intra-dataset clustering can occur when parent nodes unconstrained by the connectivity matrix are merged, resulting in the clustering of distal, nonadjacent child nodes. We note that nodes are not merged if the maximum linkage distance is exceeded. Thus, to prevent erroneous feature, the user can simply reduce their selected tolerances.

By default, alignment considers all features detected among datasets, though users may design more complicated and restrictive workflows using the DEIMoS API. For example, users may choose to only align features that appear across some number of replicates or exclude features that appear in blank samples.

To demonstrate alignment functionality, we analyzed all acquired samples (N = 912). First, we performed alignment across analytical replicates and only kept features appearing in triplicate. Next, we performed alignment across samples in both positive and negative electrospray ionization (ESI) modes. Samples were aligned by agglomerative clustering method with maximum linkage distance tolerances in each dimension of ±20 ppm, ±1.5%, and 0.3 minutes for *m/z*, drift time, and retention time, respectively (**Figure S5**).

MS2 extraction

With MS1 features of interest determined by peak detection, corresponding tandem mass spectra, if available, must be extracted and assigned to the MS1 parent ion feature. For data independent acquisition, we use non-*m/z* dimensions to assign fragments; for instance, drift time and retention time are used to match fragments in LC-IMS-MS/MS. These additional separations enable better attribution of MS2 ions to parent ions, a form of deconvolution inherent in the acquisition, but convolution artifacts can still occur.

Explicit, algorithmic deconvolution[20,25,54–56] has been implemented in DEIMoS such that MS1 and MS2 features overlapping in non-*m/z* separation dimensions are disambiguated to minimize false assignments. In this form of deconvolution, similar to the approach in Yin et al.[56], the profile(s) of non-*m/z* separation dimensions are used to identify only those ions in the MS2 with distributions that correspond to the precursor ion distribution. This technique simultaneously excludes MS2 ions arising from noise or chemical background, while also attributing MS2 ions only to precursor ions with similar separation distributions. Correspondence is determined by cosine similarity, producing a value between 0 and 100 for each separation dimension for all MS1:MS2 pairings. The user may then filter putative matches by this value, for example considering only those above some tolerance in one or more of the separation dimensions. We demonstrate the deconvolution approach in **Figure 3**.

Collision cross section calibration

To yield collision cross section (CCS) from IMS arrival time, a linear calibration must be performed using a standard tune mix containing compounds of known CCS. Drift times are reported by the instrument and calibrated against the known CCS values to yield calibration coefficients *beta* and *tfix* according to the single-field calibration equation detailed in Stow et al., and as implemented in Lee et al.[57,58] DEIMoS performs this calibration given arrival times, known CCS values, *m/z*, and nominal charge of each calibrant. Correlation coefficient and sum of residuals are reported to characterize goodness of fit. Users may also supply *beta* and *tfix* directly.

Extracted ion approach

DEIMoS can locate features based on extracted ion chromatograms (XIC), mobilograms (XIM), or multidimensional analogs. Here, a specified *m/z* of interest is supplied and the feature of maximal intensity in the remaining dimensions is returned. This technique is useful, for example, when detecting an internal standard that has been spiked into a sample or when single or mixtures of pure compounds are analyzed. We recommend multidimensional representations for targeted feature detection. An example application is included in the SI.

Isotope detection

Isotopologues, or molecules that differ only in their isotopic composition, are common in mass spectrometry analyses. In many analysis workflows, isotopologues are used to downselect the total feature list to include only the most abundant species, as well as to glean ion charge state and provide further evidence for identification by way of a detected isotopic signature. Details of the DEIMoS implementation are available in the SI.

Automation

While DEIMoS functionality is implemented as a Python API, a typical workflow has been implemented using the Snakemake workflow management system[37] and made accessible via command line interface (CLI). Users need only modify a configuration file and interact with a CLI to process input mzML files into output feature coordinates and extracted MS2. Moreover, Snakemake can automatically handle scaling to HPC and cloud resources, enabling the high-throughput processing of numerous samples. A graphical user interface (GUI) is currently in development to facilitate accessibility of DEIMoS to those without programming experience and will be reported in a subsequent manuscript.

**Results and Discussion**

In total, sample acquisition resulted in 912 data files for N samples and 10 blanks spanning 2 ionization modes (positive, negative) and 3 collision energies (10, 20, and 40 eV), cumulatively 1.1 terabytes. These data were processed by DEIMoS using feature detection, alignment, CCS calibration, and MS2 extraction by deconvolution. Where relevant, the color vision-deficient (CVD) colormap *cividis* was chosen to ensure nearly identical interpretation in CVD and non-CVD individuals, uniform perception in hue and brightness, and linearly increasing brightness[59].

Feature detection

Feature detection for a subset of a single sample has been visualized in **Figure 1**. The selected region illustrates the inherent convolution of MS1 features leads to overlap in both drift and retention time, resulting in several putative precursor ions for MS2 assignment. The requirement of explicit deconvolution to appropriately attribute ions in the MS2 spectra becomes apparent, as features are not sufficiently resolved by drift and retention time coordinates.

For all acquired samples, we compared permutations of the possible feature detection modes (3D, 2D followed by 1D, 1D followed by 2D, and iterative 1D). Notably, LC-IMS-MS/MS data exist in a 3D space; thus, the underlying features are also represented in 3D. Iterative feature detection in lower-dimensional projections simplifies the resulting data structure by summing along non-projection axes, potentially introducing artifacts. In practice, some projections, such as *m/z* versus drift time, are affected less significantly than others, such as drift time versus retention time, the latter suffering significant information loss when summing along the *m/z* axis.

We anticipated that processing the data in all 3 dimensions simultaneously would result in the greatest separation among features to better isolate the local maxima. That is, given the same feature detection tolerances across methods, 3D feature detection would theoretically afford the least overlap and, by extension, greatest number of features. Per **Figure S2**, a contrary result was thus surprising. However, the coordinates of the features detected by lower dimensional projections are not congruent with the 3D approach (**Figure S3**). This signals that the projections along the various data axes, whether 1D or 2D, aggregate signal to the point of, in some cases, losing the underlying feature defined in 3D – the sum operation along a given axis "merges" previously separated features, skewing the coordinate in that dimension.

However, this phenomenon is pronounced to varying degree among methods: the most comparable technique, as implemented by MS-DIAL – *m/z* versus RT followed by DT – results in poor agreement when considering strict tolerances (only ~13% intersection in both positive and negative mode), but intersection increases substantially (to ~90%) when imposing the same tolerances that would be used in cross-sample alignment. That is, tolerances that would result in the combining of those features anyway. In this case, the lower dimensional projections result in slight differences in feature coordinates, but in practical application would be treated as "same". The difference could result in less accurate characterization of feature coordinates, for example exact *m/z*, CCS calculations from drift time, and/or retention time. Critically, the order of peak picking operations had a large impact on the number and composition of the features detected (e.g., comparing 1D-2D to 2D-1D operations, as well as the consecutive 1D operations).

A key advantage of feature detection in native dimensionality is that computation time does not scale with the number of features (**Figure S2**), which is further discussed in the SI.

Alignment

We selected example datasets to highlight both S-shaped (**Figure 2**) and linear (**Figure S4**) relationships in pre-alignment retention time and illustrate the flexibility of the SVR-based approach. In these data, *m/z* and drift time were already sufficiently aligned; as such, plots for *m/z* and drift time were omitted. The potential difference realized by kernel selection motivates visual confirmation of the alignment relationship between samples: an RBF kernel applied to linearly related samples would be considered overfit, whereas a linear kernel applied to a nonlinear case would achieve poor alignment. While SVR was selected here for its broad applicability to both linear and nonlinear alignment, many approaches have been successfully developed in this space[2,13,22,53,60–62], and SVR is not necessarily superior.

Subsequent cross-sample alignment by agglomerative clustering resulted in 11,698 and 14,784 features for positive and negative mode appearing across each of 3 analytical replicates, excluding features that appeared in all blank samples. Tolerances of ±20 ppm in *m/z*, ±1.5% in drift time, and ±0.3 minutes in retention time were selected according to expected variance across samples. However, as demonstrated in **Figure S5,** agglomerative clustering resulted in tighter linkages, indicating inter-sample variance is lower than typical peak width. This variance reaches near-maximal values at 15 ppm in *m/z*, 1% in drift time, and 0.2 minutes in retention. Thus, exploratory analysis of the data enables users to characterize inter-feature and inter-sample variance to set appropriate tolerances relative to the source of maximum variance, though agglomerative clustering is relatively forgiving so long as tolerances are not too small.

<u>MS2 extraction</u>

For the MS1 features shown in **Figure 1**, we performed deconvolution to putatively assign ions in the MS2 spectra to corresponding ions in the MS1 spectra. The utility of deconvolution is highlighted in **Figure 3**. A window-based approach yields MS2 spectra that, in the worst case, erroneously include all ions for all MS1 precursors to give identical MS2 spectra or, in the best case, results in only two distinct spectra among precursors. The limitation here is visualized by the plot of drift time versus retention time, where only two overlapping peaks emerge. Naïve assignment would thus yield convolved spectra, the degree of convolution depending on window selection.

Notably, deconvolution results may differ substantially if employing the cosine similarity scores of drift versus retention time, as false positives can occur if using retention time alone. For example, the masses between 77 Da and 83 Da would be assigned to the precursor with *m/z* 212 using retention time, whereas they are sufficiently separated in drift time (**Figure 3**).

## Conclusion

Metabolomics and exposomics data processing tools offer immense value for diagnosis of disease, evaluation of environmental exposures, and discovery of novel molecules. However, few open-source solutions are currently positioned to fully leverage the latest instrumentation. Importantly, though demonstrated for LC-IMS-MS/MS data, DEIMoS's architecture supports extension to other measurement modalities, such as cryogenic infrared spectroscopy, minimizing development barriers as instrumentation evolves. Further, all development has been accomplished using design principles necessary for the long-term success for metabolomics data: format interoperability, workflow flexibility, open-source software implementation, community development, and reproducibility.

## Acknowledgments

This research was supported by the National Institutes of Health, National Institute of Environmental Health Sciences grant U2CES030170. Additional support was provided by the Pacific Northwest National Laboratory (PNNL), Laboratory Directed Research and Development Program, and is a contribution of the Biomedical Resilience & Readiness in Adverse Operating Environments (BRAVE) Agile project. The authors wish to thank Mr. Nathan Johnson of PNNL for assistance with graphics. LC-IMS-MS/MS measurements were performed in the Environmental Molecular Sciences Laboratory, a national scientific user facility sponsored by the U.S. Department of Energy Office of Biological and Environmental Research and located on the campus of Pacific Northwest National Laboratory (PNNL) in Richland, Washington. PNNL is a multi-program national laboratory operated by Battelle for the DOE under Contract DE-AC05-76RLO 1830.

## References

1. Wang, P., Yang, P., Arthur, J. & Yang, J. Y. H. A dynamic wavelet-based algorithm for pre-processing tandem mass spectrometry data. *Bioinformatics* **26**, 2242–2249 (2010).
2. Katajamaa, M. & Orešič, M. Data processing for mass spectrometry-based metabolomics. *J. Chromatogr. A* **1158**, 318–328 (2007).
3. Kiefer, P., Schmitt, U. & Vorholt, J. A. eMZed: an open source framework in Python for rapid and interactive development of LC/MS data analysis workflows. *Bioinformatics* **29**, 963–964 (2013).
4. Tautenhahn, R., Böttcher, C. & Neumann, S. Highly sensitive feature detection for high resolution LC/MS. *BMC Bioinformatics* **9**, 504 (2008).
5. Du, P., Kibbe, W. A. & Lin, S. M. Improved peak detection in mass spectrum by incorporating continuous wavelet transform-based pattern matching. *Bioinformatics* **22**, 2059–2065 (2006).
6. DeFelice, B. C. *et al.* Mass Spectral Feature List Optimizer (MS-FLO): A Tool To Minimize False Positive Peak Reports in Untargeted Liquid Chromatography–Mass Spectroscopy (LC-MS) Data Processing. *Anal. Chem.* **89**, 3250–3255 (2017).
7. Broeckling, C. D., Reddy, I. R., Duran, A. L., Zhao, X. & Sumner, L. W. MET-IDEA:  Data Extraction Tool for Mass Spectrometry-Based Metabolomics. *Anal. Chem.* **78**, 4334–4341 (2006).
8. Pluskal, T., Castillo, S., Villar-Briones, A. & Orešič, M. MZmine 2: Modular framework for processing, visualizing, and analyzing mass spectrometry-based molecular profile data. *BMC Bioinformatics* **11**, 395 (2010).
9. Hastings, C. A., Norton, S. M. & Roy, S. New algorithms for processing and peak detection in liquid chromatography/mass spectrometry data. *Rapid Commun. Mass Spectrom.* **16**, 462–467 (2002).
10. Röst, H. L. *et al.* OpenMS: a flexible open-source software platform for mass spectrometry data analysis. *Nat. Methods* **13**, 741–748 (2016).
11. Monroe, M. E. *et al.* VIPER: an advanced software package to support high-throughput LC-MS peptide identification. *Bioinformatics* **23**, 2021–2023 (2007).
12. French, W. R. *et al.* Wavelet-Based Peak Detection and a New Charge Inference Procedure for MS/MS Implemented

in ProteoWizard's msConvert. *J. Proteome Res.* **14**, 1299–1307 (2015).

13. Smith, C. A., Want, E. J., O'Maille, G., Abagyan, R. & Siuzdak, G. XCMS: Processing Mass Spectrometry Data for Metabolite Profiling Using Nonlinear Peak Alignment, Matching, and Identification. *Anal. Chem.* **78**, 779–787 (2006).
14. Blaženović, I. *et al.* Increasing Compound Identification Rates in Untargeted Lipidomics Research with Liquid Chromatography Drift Time–Ion Mobility Mass Spectrometry. *Anal. Chem.* **90**, 10758–10764 (2018).
15. Dodds, J. N. & Baker, E. S. Ion Mobility Spectrometry: Fundamental Concepts, Instrumentation, Applications, and the Road Ahead. *J. Am. Soc. Mass Spectrom.* **30**, 2185–2195 (2019).
16. Lanucara, F., Holman, S. W., Gray, C. J. & Eyers, C. E. The power of ion mobility-mass spectrometry for structural characterization and the study of conformational dynamics. *Nat. Chem.* **6**, 281–294 (2014).
17. May, J. C. & McLean, J. A. Ion mobility-mass spectrometry: time-dispersive instrumentation. *Anal. Chem.* **87**, 1422–1436 (2015).
18. Metz, T. O. *et al.* Integrating ion mobility spectrometry into mass spectrometry-based exposome measurements: what can it add and how far can it go? *Bioanalysis* **9**, 81–98 (2017).
19. Paglia, G. *et al.* Ion mobility derived collision cross sections to support metabolomics applications. *Anal. Chem.* **86**, 3985–3993 (2014).
20. Tsugawa, H. *et al.* MS-DIAL 4: accelerating lipidomics using an MS/MS, CCS, and retention time atlas. *bioRxiv* 2020.02.11.944900 (2020) doi:10.1101/2020.02.11.944900.
21. Crowell, K. L. *et al.* LC-IMS-MS Feature Finder: detecting multidimensional liquid chromatography, ion mobility and mass spectrometry features in complex datasets. *Bioinformatics* **29**, 2804–2805 (2013).
22. Fernández-Ochoa, Á. *et al.* A Case Report of Switching from Specific Vendor-Based to R-Based Pipelines for Untargeted LC-MS Metabolomics. *Metabolites* **10**, 28 (2020).
23. Aiche, S. *et al.* Workflows for automated downstream data analysis and visualization in large-scale computational mass spectrometry. *PROTEOMICS* **15**, 1443–1447 (2015).
24. Kessner, D., Chambers, M., Burke, R., Agus, D. & Mallick, P. ProteoWizard: open source software for rapid proteomics tools development. *Bioinformatics* **24**, 2534–2536 (2008).
25. Benton, H. P., Wong, D. M., Trauger, S. A. & Siuzdak, G. XCMS2: Processing Tandem Mass Spectrometry Data for Metabolite Identification and Structural Characterization. *Anal. Chem.* **80**, 6382–6389 (2008).
26. Melamud, E., Vastag, L. & Rabinowitz, J. D. Metabolomic Analysis and Visualization Engine for LC−MS Data. *Anal. Chem.* **82**, 9818–9826 (2010).
27. Chong, J. *et al.* MetaboAnalyst 4.0: towards more transparent and integrative metabolomics analysis. *Nucleic Acids Res.* **46**, W486–W494 (2018).
28. Harris, C. R. *et al.* Array programming with NumPy. *Nature* **585**, 357–362 (2020).
29. Virtanen, P. *et al.* SciPy 1.0: fundamental algorithms for scientific computing in Python. *Nat. Methods* **17**, 261–272 (2020).
30. Zhang, X. *et al.* SPE-IMS-MS: An automated platform for sub-sixty second surveillance of endogenous metabolites and xenobiotics in biofluids. *Clin. Mass Spectrom. Mar Calif* **2**, 1–10 (2016).
31. Warnke, S., Faleh, A. B., Pellegrinelli, R. P., Yalovenko, N. & Rizzo, T. R. Combining ultra-high resolution ion mobility spectrometry with cryogenic IR spectroscopy for the study of biomolecular ions. *Faraday Discuss.* **217**, 114–125 (2019).
32. Deng, L. *et al.* Ultra-High Resolution Ion Mobility Separations Utilizing Traveling Waves in a 13 m Serpentine Path Length Structures for Lossless Ion Manipulations Module. *Anal. Chem.* **88**, 8957–8964 (2016).
33. Simón-Manso, Y. *et al.* Metabolite Profiling of a NIST Standard Reference Material for Human Plasma (SRM 1950): GC-MS, LC-MS, NMR, and Clinical Laboratory Analyses, Libraries, and Web-Based Resources. https://pubs.acs.org/doi/pdf/10.1021/ac402503m (2013) doi:10.1021/ac402503m.
34. Matyash, V., Liebisch, G., Kurzchalia, T. V., Shevchenko, A. & Schwudke, D. Lipid extraction by methyl-tert-butyl ether for high-throughput lipidomics*s. *J. Lipid Res.* **49**, 1137–1146 (2008).
35. Lee, D. Y., Kind, T., Yoon, Y.-R., Fiehn, O. & Liu, K.-H. Comparative evaluation of extraction methods for simultaneous mass-spectrometric analysis of complex lipids and primary metabolites from human blood plasma. *Anal. Bioanal. Chem.* **406**, 7275–7286 (2014).
36. Pittard, W. S. & Li, S. The Essential Toolbox of Data Science: Python, R, Git, and Docker. in *Computational Methods and Data Analysis for Metabolomics* (ed. Li, S.) 265–311 (Springer US, 2020). doi:10.1007/978-1-0716-0239-3_15.
37. Köster, J. & Rahmann, S. Snakemake—a scalable bioinformatics workflow engine. *Bioinformatics* **28**, 2520–2522 (2012).
38. Chang, H.-Y. *et al.* A Practical Guide to Metabolomics Software Development. *Anal. Chem.* **93**, 1912–1923 (2021).
39. Anaconda. *Anaconda* https://www.anaconda.com/.
40. PyPI. *PyPI* https://pypi.org/.
41. Sphinx. https://www.sphinx-doc.org/en/master/.
42. pytest. https://docs.pytest.org/en/stable/.
43. Git. https://git-scm.com/.
44. Martens, L. *et al.* mzML—a Community Standard for Mass Spectrometry Data. *Mol. Cell. Proteomics* **10**, (2011).
45. Jeff Reback *et al. pandas*. (Zenodo, 2020). doi:10.5281/zenodo.3715232.
46. McKinney, W. Data Structures for Statistical Computing in Python. in 56–61 (2010). doi:10.25080/Majora-92bf1922-00a.
47. HDF5. *The HDF Group* https://www.hdfgroup.org/downloads/hdf5/.

48. Nuñez, J. R. *et al.* Evaluation of In Silico Multifeature Libraries for Providing Evidence for the Presence of Small Molecules in Synthetic Blinded Samples. *J. Chem. Inf. Model.* **59**, 4052–4060 (2019).
49. Kyle, J. E. *et al.* LIQUID: an-open source software for identifying lipids in LC-MS/MS-based lipidomics data. *Bioinformatics* **33**, 1744–1746 (2017).
50. Wang, M. *et al.* Sharing and community curation of mass spectrometry data with Global Natural Products Social Molecular Networking. *Nat. Biotechnol.* **34**, 828–837 (2016).
51. Hussong, R., Tholey, A. & Hildebrandt, A. Efficient Analysis of Mass Spectrometry Data Using the Isotope Wavelet. *AIP Conf. Proc.* **940**, 139–149 (2007).
52. Pedregosa, F. *et al.* Scikit-learn: Machine Learning in Python. *J. Mach. Learn. Res.* **12**, 2825–2830 (2011).
53. Watrous, J. D. *et al.* Visualization, Quantification, and Alignment of Spectral Drift in Population Scale Untargeted Metabolomics Data. *Anal. Chem.* **89**, 1399–1404 (2017).
54. Smirnov, A. *et al.* ADAP-GC 4.0: Application of Clustering-Assisted Multivariate Curve Resolution to Spectral Deconvolution of Gas Chromatography–Mass Spectrometry Metabolomics Data. *Anal. Chem.* **91**, 9069–9077 (2019).
55. Tada, I. *et al.* Correlation-Based Deconvolution (CorrDec) To Generate High-Quality MS2 Spectra from Data-Independent Acquisition in Multisample Studies. *Anal. Chem.* **92**, 11310–11317 (2020).
56. Yin, Y., Wang, R., Cai, Y., Wang, Z. & Zhu, Z.-J. DecoMetDIA: Deconvolution of Multiplexed MS/MS Spectra for Metabolite Identification in SWATH-MS-Based Untargeted Metabolomics. *Anal. Chem.* **91**, 11897–11904 (2019).
57. Stow, S. M. *et al.* An Interlaboratory Evaluation of Drift Tube Ion Mobility-Mass Spectrometry Collision Cross Section Measurements. *Anal. Chem.* **89**, 9048–9055 (2017).
58. Lee, J.-Y. *et al.* AutoCCS: automated collision cross-section calculation software for ion mobility spectrometry–mass spectrometry. *Bioinformatics* (2021) doi:10.1093/bioinformatics/btab429.
59. Nuñez, J. R., Anderton, C. R. & Renslow, R. S. Optimizing colormaps with consideration for color vision deficiency to enable accurate interpretation of scientific data. *PLOS ONE* **13**, e0199239 (2018).
60. Gupta, S., Ahadi, S., Zhou, W. & Röst, H. DIAlignR Provides Precise Retention Time Alignment Across Distant Runs in DIA and Targeted Proteomics. *Mol. Cell. Proteomics* **18**, 806–817 (2019).
61. Nielsen, N.-P. V., Carstensen, J. M. & Smedsgaard, J. Aligning of single and multiple wavelength chromatographic profiles for chemometric data analysis using correlation optimised warping. *J. Chromatogr. A* **805**, 17–35 (1998).
62. Prince, J. T. & Marcotte, E. M. Chromatographic Alignment of ESI-LC-MS Proteomics Data Sets by Ordered Bijective Interpolated Warping. *Anal. Chem.* **78**, 6140–6152 (2006).

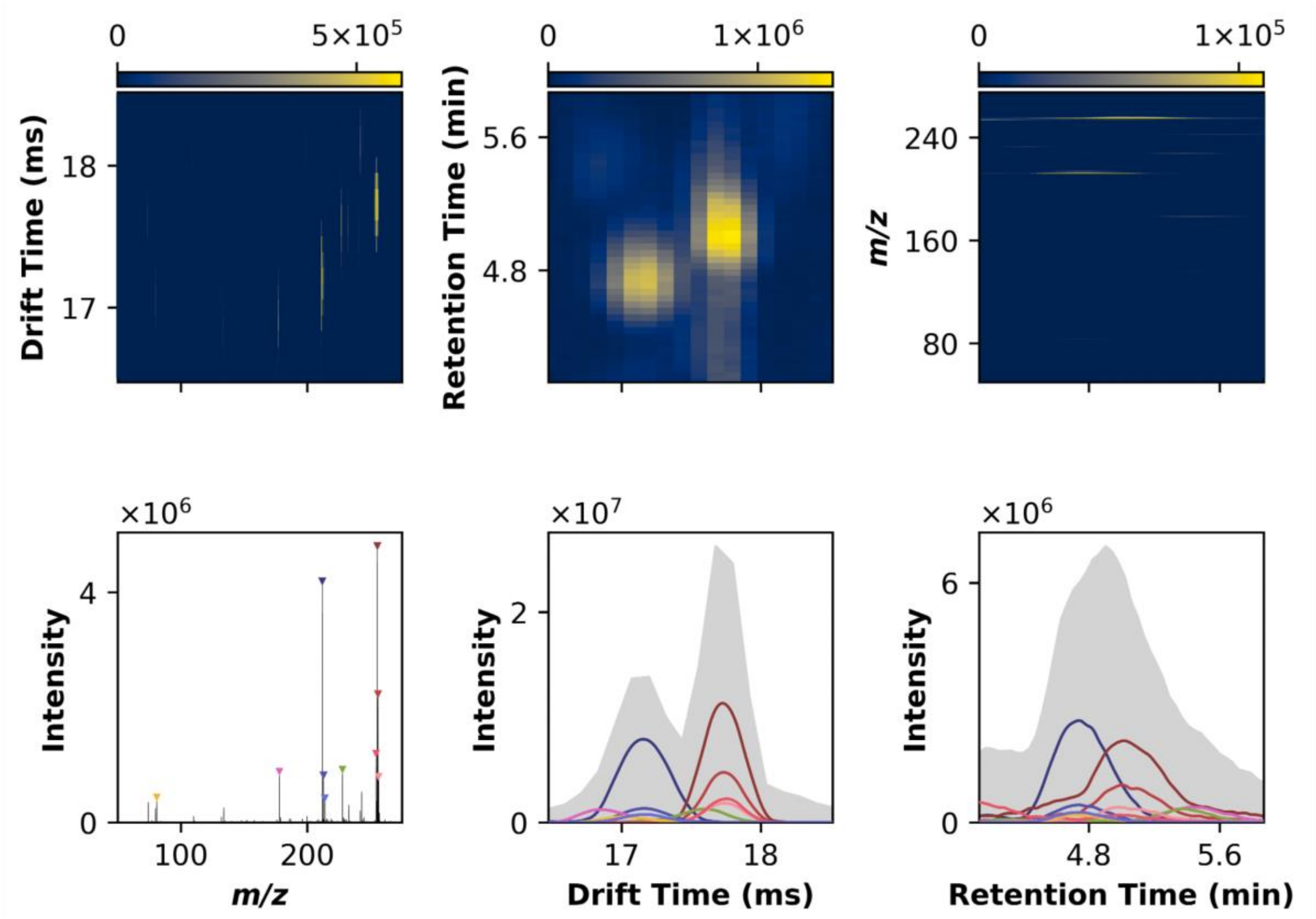

**Figure 1. Multidimensional peak detection.** Peak detection involves convolving the input signal in N dimensions (here, in LC-IMS-MS, 3D) with a maximum filter. The input and maximum filtered arrays are then compared point-by-point and, where equal, a local maximum is indicated. While the data is collected in 3D, this approach is best visualized in 2D and 1D projections, capturing all lower dimensional representations of the underlying 3D data. Note a well-defined peak in a given 2D view may or may not correspond to a true 3D apex, or can be the product of multiple underlying features. It is thus important to interpret the 1D projections carefully. For this subset of the data, the top 10 most intense local maxima are shown, colored by *m/z*, with similar *m/z* (i.e. isopologues) sharing hues.

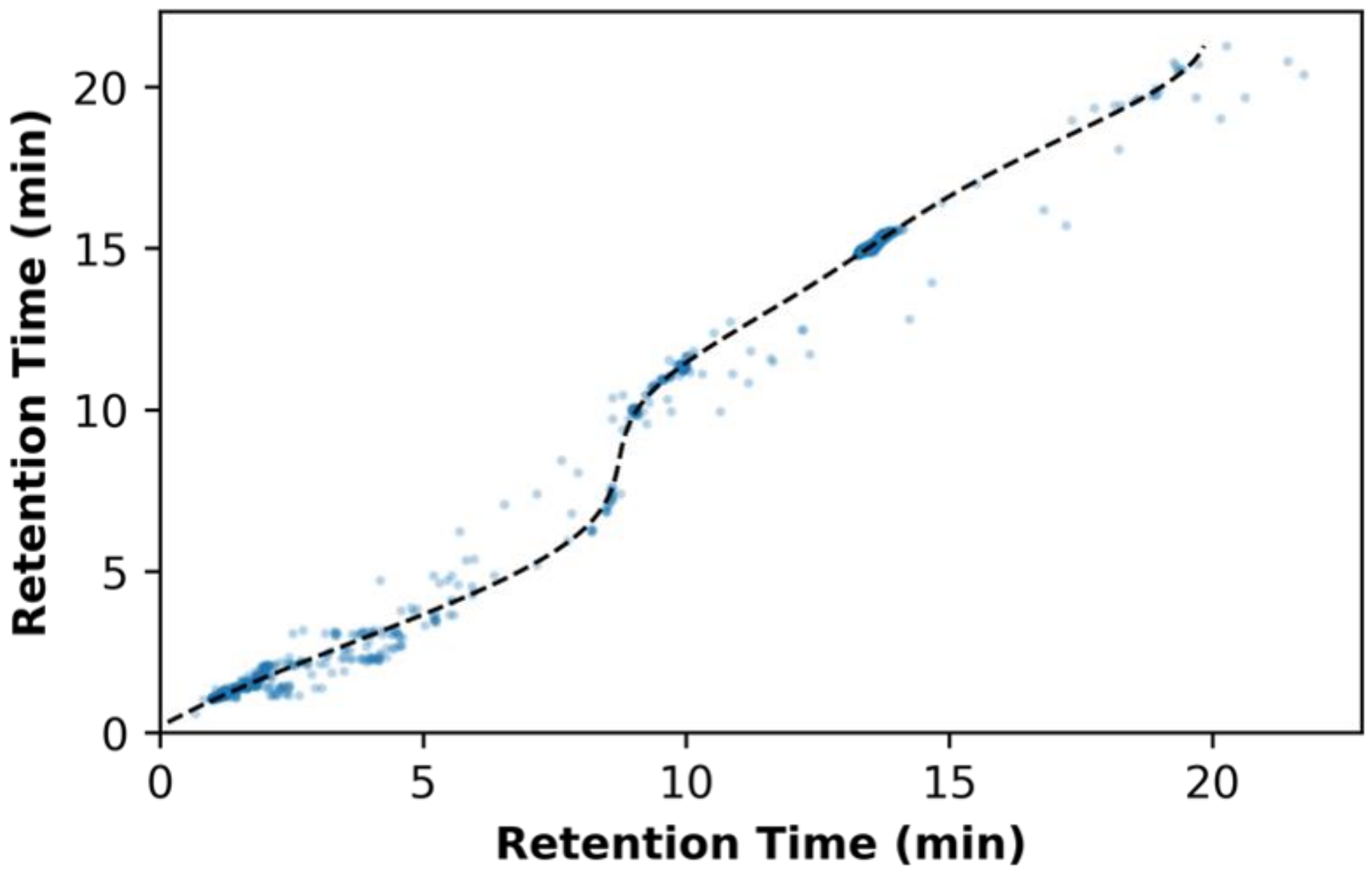

**Figure 2. Nonlinear alignment by support vector regression**. Support vector regression (SVR) was evaluated here on the retention time dimension between 2 illustrative samples described by a nonlinear, "S-shaped" relationship in retention time. To model this relationship, a radial basis function (RBF) kernel was selected. Measurements between samples varied negligibly in drift time and *m/z*, and thus alignment was only necessary in retention time. An example involving samples with a linear relationship in retention time is included in **Figure S4**.

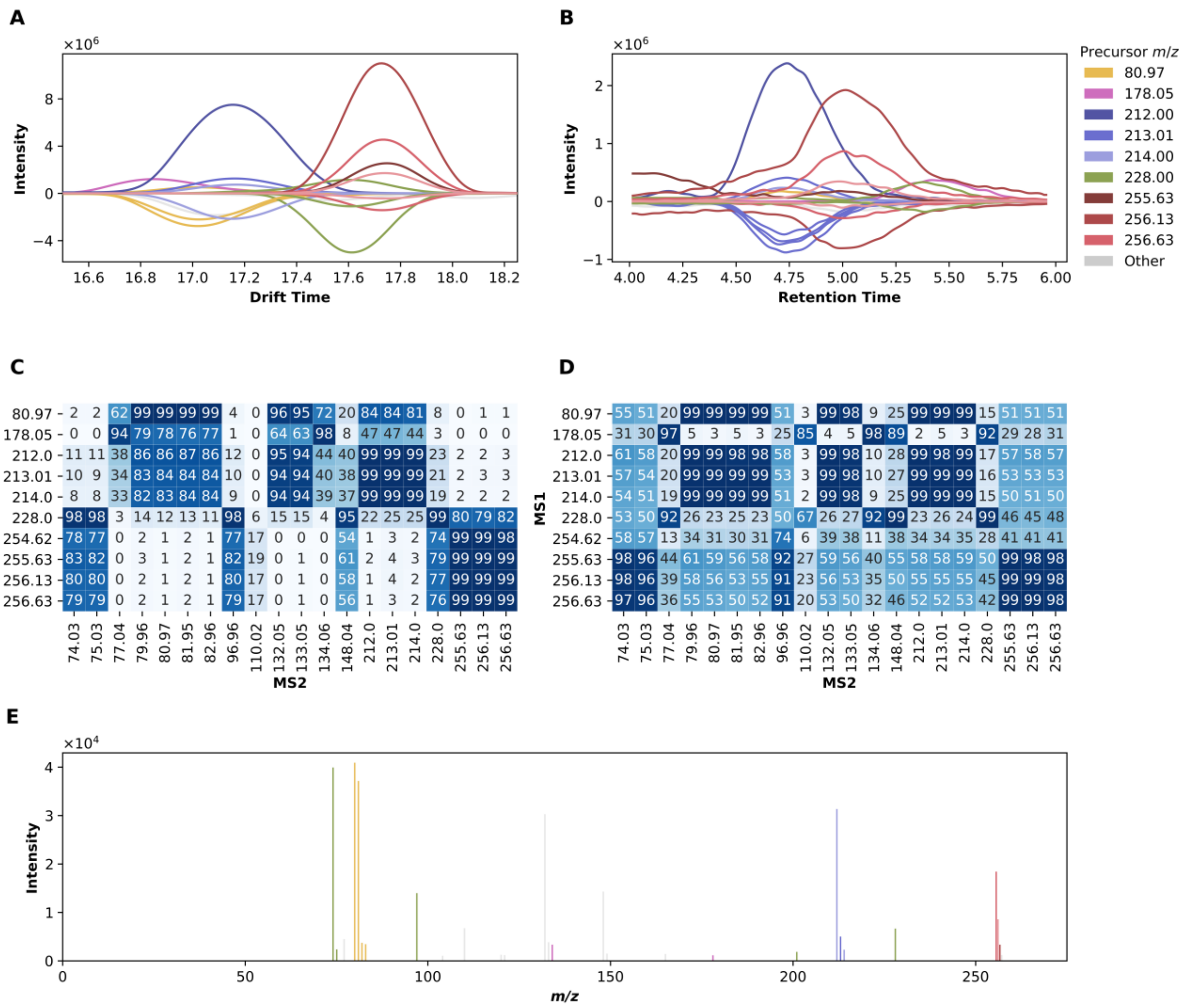

**Figure 3. MS2 deconvolution.** The MS2 spectra belonging to the MS1 features highlighted in **Figure 1** are algorithmically deconvolved. The profiles of the MS1 features are indicated by respective colors, plotted along the positive y-axis for drift time (A) and retention time (B). These are accompanied by corresponding MS2 profiles, plotted along the negative y-axis, colored according to the closest matching MS1 profile by cosine similarity. Panels C and D show the pairwise cosine similarity of MS1 and MS2 profiles for drift and retention time, respectively. In panel E, ions in the MS2 spectra are colored according to the closest matching MS1 drift time profile. As in **Figure 1**, only the 10 most intense MS1 features were explicitly colored; the ions in the MS2 spectra corresponding to remaining MS1 features, or not sufficiently similar to an MS1 precursor, are indicated in gray. Note that disambiguating among isotopologues is not possible here, thus the difference in color among isotopic groupings is largely superficial.

## For Table of Contents Only

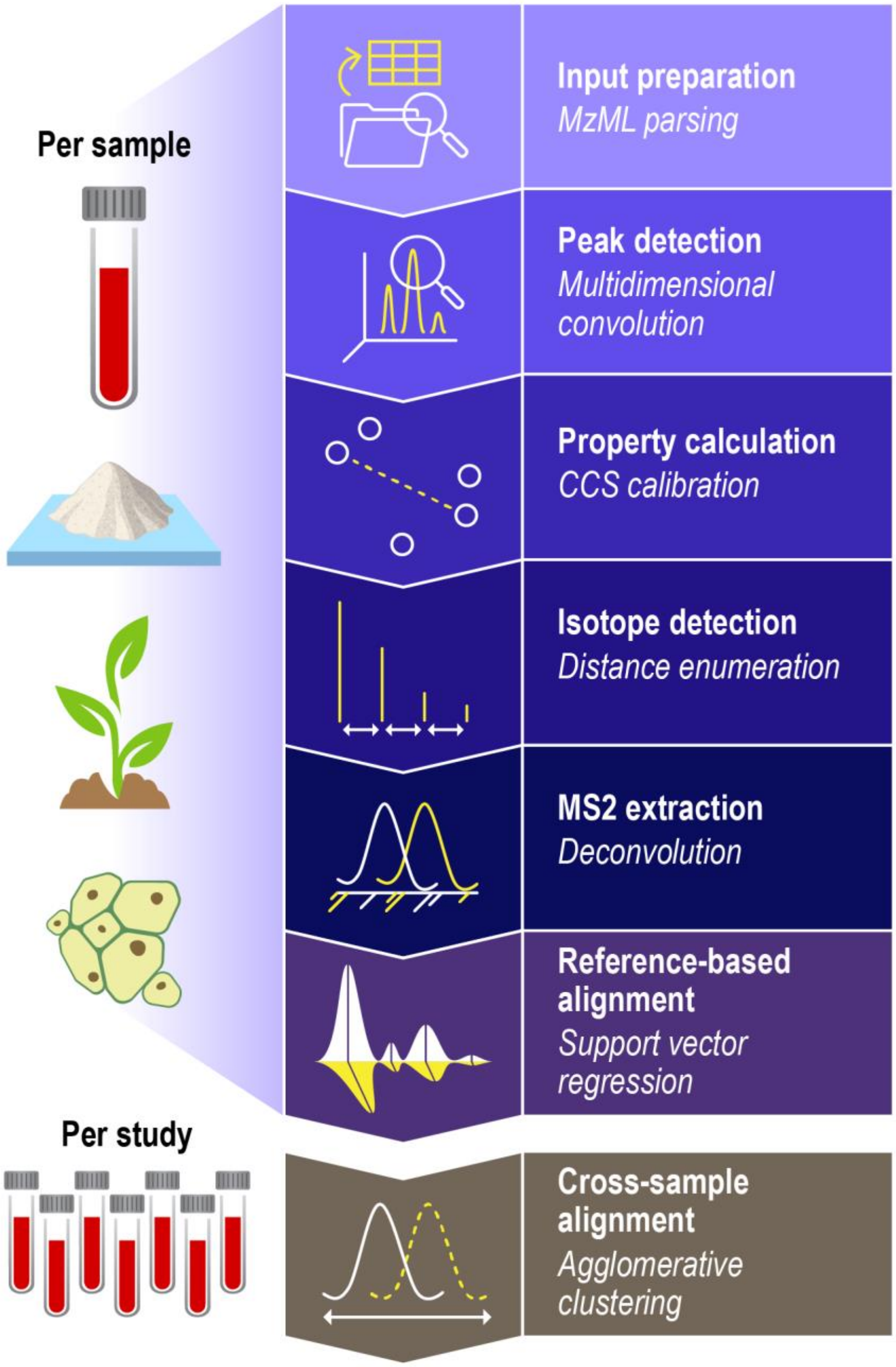

**Overview of DEIMoS functionality.** High level overview of available DEIMoS functionality, with operations delineated as "per sample" versus "per study". That is, the former operations are performed for each instrument acquisition, whereas the latter is performed among all comparable samples acquired. Functionality and underlying methods are easily extended or modified.